\documentclass[aps,longbibliography,superscriptaddress,reprint,amsmath,amssymb]{revtex4-2}
\usepackage[colorlinks=true,urlcolor=blue,citecolor=blue,linkcolor=blue]{hyperref} 
\usepackage{graphicx}
\usepackage{dcolumn}
\usepackage{bm}

\usepackage{subcaption} 
\usepackage{overpic}   
\usepackage{braket}
\usepackage{float}
\usepackage{color}
\usepackage{soul}
\usepackage{xcolor}

\makeatletter

\newcommand{\Rmnum}[1]{\expandafter\@slowromancap\romannumeral #1@}
\makeatother

\begin{document}

\preprint{APS/123-QED}

\title{Experimental Proposal on Non-Abelian Aharonov-Bohm Caging Effect \\ with a Single Trapped Ion}

\author{Zhiyuan Liu}\email{These authors contribute equally.}
\affiliation{CAS Key Laboratory of Microscale Magnetic Resonance and School of Physical Sciences, University of Science and Technology of China, Hefei 230026, China}
\affiliation{Anhui Province Key Laboratory of Scientific Instrument Development and Application, University of Science and Technology of China, Hefei 230026, China}

\author{Wanchao Yao}\email{These authors contribute equally.}
\affiliation{CAS Key Laboratory of Microscale Magnetic Resonance and School of Physical Sciences, University of Science and Technology of China, Hefei 230026, China}
\affiliation{Anhui Province Key Laboratory of Scientific Instrument Development and Application, University of Science and Technology of China, Hefei 230026, China}

\author{Sai Li}
\affiliation{Key Laboratory of Atomic and Subatomic Structure and Quantum Control (Ministry of Education), Guangdong Basic Research Center of Excellence for Structure and Fundamental Interactions of Matter, and School of Physics, South China Normal University, Guangzhou 510006, China}
\affiliation{Guangdong Provincial Key Laboratory of Quantum Engineering and Quantum Materials, Guangdong-Hong Kong Joint Laboratory of Quantum Matter, and Frontier Research Institute for Physics, South China Normal University, Guangzhou 510006, China}

\author{Yi Li}
\affiliation{CAS Key Laboratory of Microscale Magnetic Resonance and School of Physical Sciences, University of Science and Technology of China, Hefei 230026, China}
\affiliation{Anhui Province Key Laboratory of Scientific Instrument Development and Application, University of Science and Technology of China, Hefei 230026, China}
\affiliation{Hefei National Laboratory, University of Science and Technology of China, Hefei 230088, China}

\author{Yue Li}
\affiliation{CAS Key Laboratory of Microscale Magnetic Resonance and School of Physical Sciences, University of Science and Technology of China, Hefei 230026, China}
\affiliation{Anhui Province Key Laboratory of Scientific Instrument Development and Application, University of Science and Technology of China, Hefei 230026, China}

\author{Zheng-Yuan Xue}\email{zyxue@scnu.edu.cn}
\affiliation{Key Laboratory of Atomic and Subatomic Structure and Quantum Control (Ministry of Education), Guangdong Basic Research Center of Excellence for Structure and Fundamental Interactions of Matter, and School of Physics, South China Normal University, Guangzhou 510006, China}
\affiliation{Guangdong Provincial Key Laboratory of Quantum Engineering and Quantum Materials, Guangdong-Hong Kong Joint Laboratory of Quantum Matter, and Frontier Research Institute for Physics, South China Normal University, Guangzhou 510006, China}
\affiliation{Hefei National Laboratory, Hefei 230088, China}

\author{Yiheng Lin} \email{yiheng@ustc.edu.cn}
\affiliation{CAS Key Laboratory of Microscale Magnetic Resonance and School of Physical Sciences, University of Science and Technology of China, Hefei 230026, China}
\affiliation{Anhui Province Key Laboratory of Scientific Instrument Development and Application, University of Science and Technology of China, Hefei 230026, China}
\affiliation{Hefei National Laboratory, University of Science and Technology of China, Hefei 230088, China}


\begin{abstract}
In the lattice system, when the synthetic flux reaches a $\pi$ phase along a closed loop under the synthetic gauge field, destructive interference occurs and gives rise to the localization phenomenon. This is known as the Aharonov-Bohm (AB) caging effect. It provides a powerful tool for the study of quantum transportation and dynamical effects. In the system where lattice sites possess internal structure and the underlying gauge field is non-Abelian, localization can also occur, forming the non-Abelian AB caging. Here, we propose an experimental scheme to synthesize non-Abelian gauge fields with a single trapped ion by coupling multiple internal levels and Fock states in its motion via laser fields. In contrast to the Abelian AB caging, we numerically observe that the non-Abelian AB caging occurs either when the interference matrix is nilpotent, or when the initial state is specifically set. Our experimental scheme broadens the study of localization phenomena and provides a novel tool for the study of non-Abelian physics.
\end{abstract}

                              
\maketitle


\begin{figure*}[ht]
    \centering
    \begin{minipage}{0.4\textwidth}
        \begin{subfigure}{\textwidth}
            \begin{overpic}[width=\textwidth]{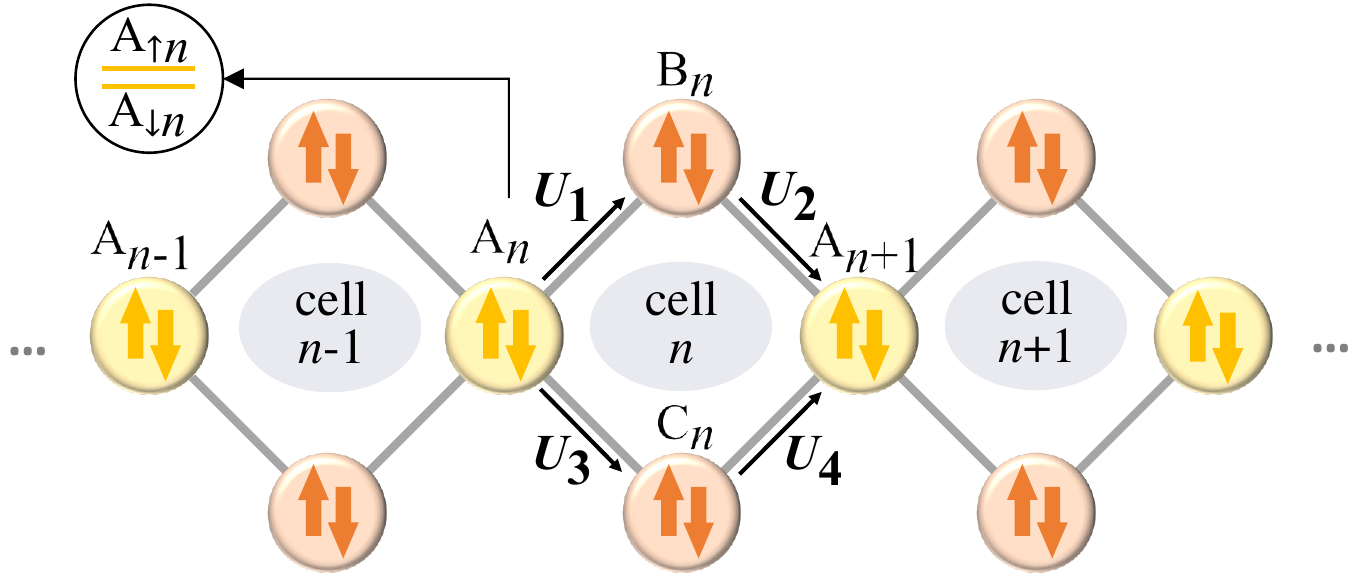}
                \put(0,43){(a)}
            \end{overpic}
            \label{fig:1_subfig_a}
        \end{subfigure}
        \begin{subfigure}{\textwidth}
            \begin{overpic}[width=\textwidth]{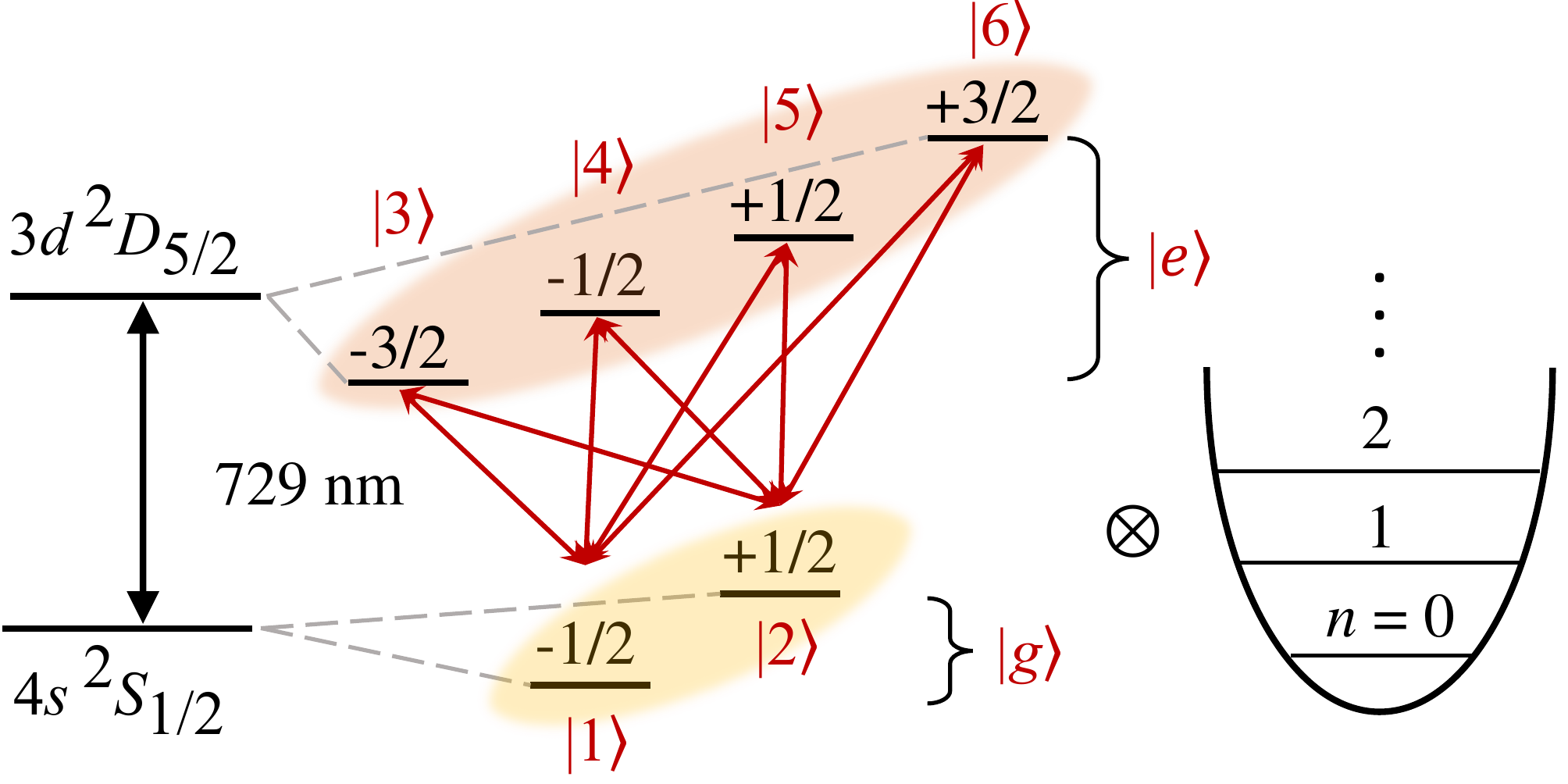}
                \put(0,43){(b)}
            \end{overpic}
            \label{fig:1_subfig_b}
        \end{subfigure}
    \end{minipage}%
    \hfill
    \begin{minipage}{0.6\textwidth}
        \begin{subfigure}{\textwidth}
            \begin{overpic}[width=\textwidth]{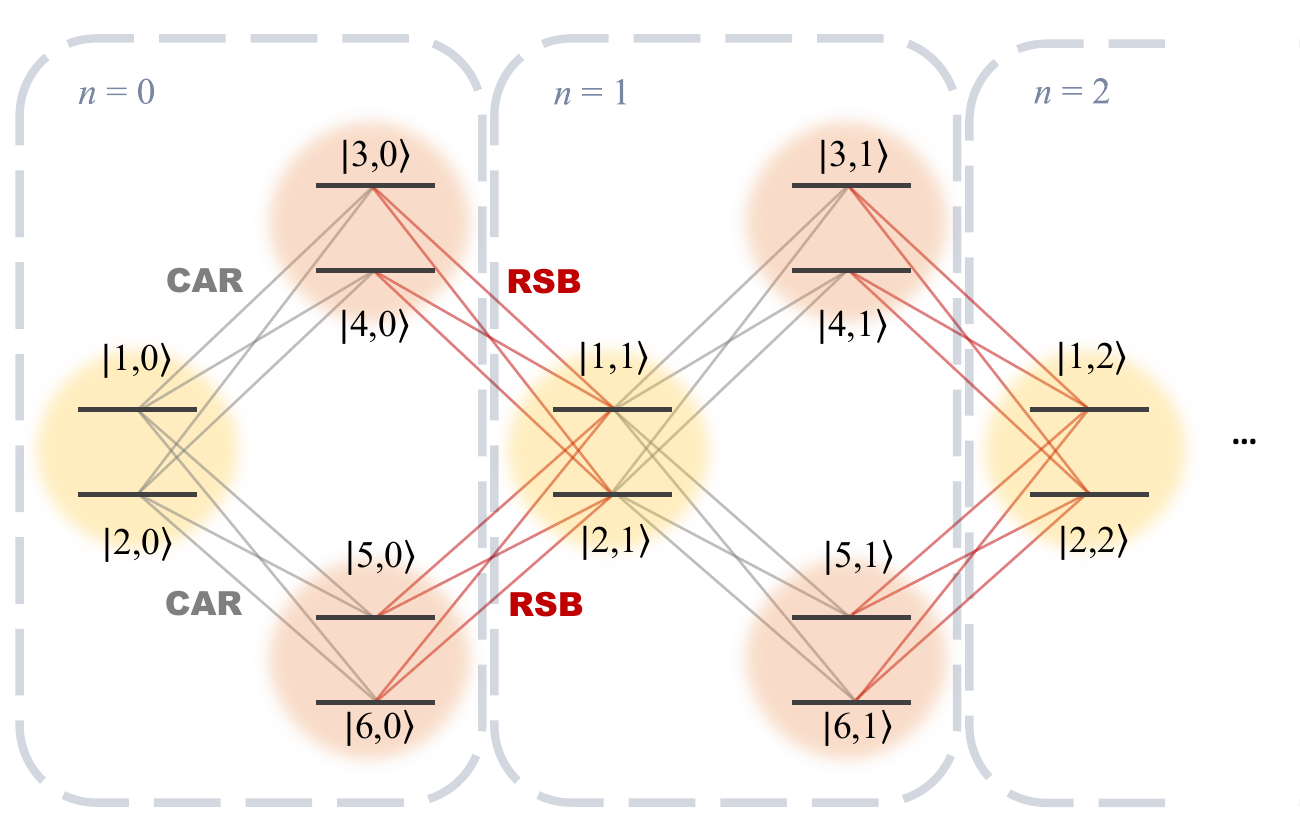}
                \put(0,66){(c)}
            \end{overpic}
            \label{fig:1_subfig_c}
        \end{subfigure}
    \end{minipage}
    \caption{(a) Schematic illustration of a one-dimensional (1D) periodic rhombic lattice. The lattice is composed of unit cells, each of which consists of three sites labeled as A, B, and C. Every site accommodates two spin modes. (b) Internal level scheme of the \(^{40}\mathrm{Ca}^{+}\) ion with Fock states in its motion. Each transition between \(S_{1/2}\) and \(D_{5/2}\) is accessible using a single narrowband laser at 729 nm. (c) Encoding protocol utilizing the internal and motional states of the \(^{40}\mathrm{Ca}^{+}\) ion to realize the 1D lattice, utilizing CAR and RSB transitions to create the U(2) gauge field. Product state \(\ket{g}\otimes\ket{n}\ (\ket{e}\otimes\ket{n})\) are shown as \(\ket{g,n}\ (\ket{e,n})\) for simplicity.}
    \label{fig:1}
\end{figure*}

In the lattice system, recent advances in quantum simulation develop the ability to synthesize a highly tunable gauge field, which facilitates the exploration of a series of novel phenomena \cite{liu25, neef23, shan24, 3dyudapeng, cpl-synthetic}. One of the profound results is the Aharonov-Bohm (AB) caging effect \cite{original1}. When the synthetic flux reaches a $\pi$ phase along a closed loop, destructive interference prevents the spread of the wave function, causing the localization phenomenon. The AB caging has been realized on various platforms, including trapped ion \cite{ion2011}, ultracold atom \cite{yanbo1}, photonic waveguide \cite{guangbodao1} and Rydberg atom \cite{chen2024strongly}. It has provided an important tool for the study of quantum transportation and dynamical effects \cite{guangbodao2, fang12, noguchi14, kawaguchi24, liberto19, gligoric19}. However, AB caging has been largely restricted to the system where the lattice sites have no internal structure, and the underlying gauge field is Abelian \cite{3dyudapeng, chen2024strongly}. When the lattice sites possess internal degrees of freedom, the dimension of the system becomes higher, and the gauge field can turn into non-Abelian \cite{chiral}. The particle in the lattice may transport to any of the internal levels in its neighbor, raising the difficulty for the non-Abelian AB caging. It also poses challenges to the experimental manipulation of quantum states. 

The non-Abelian AB caging is predicted to host remarkably different characteristics compared with the Abelian one \cite{PRA}. In a recent study, the non-Abelian AB caging was realized on the topolectrical circuit, where the influence of disorder was studied \cite{dianlu}. However, the experimental realization of non-Abelian AB caging still lacks on quantum simulator platforms. Recently, experimental proposal on the quantum electrodynamics system has been proposed \cite{PRA}. On the other hand, trapped ion platform has the advantage of long coherence time and high fidelity quantum operations \cite{annual}. By utilizing multiple internal levels, also known as the qudit, and the Fock states in its motion \cite{ringbauer2022, meth2025simulating}, a complex lattice structure can be achieved. Flexible and controllable connections can be realized via the amplitude and phase of the laser fields. These features make trapped ions a leading platform for quantum simulation and suitable for the study of complicated synthetic gauge fields \cite{ion2011, ionABring, ann96, ann97, ann76, ann98, meth2025simulating, ann100, wang2024,hou2024}. 

In this Letter, we propose an experimental scheme for observing the non-Abelian AB caging effect with a single trapped ion. We encode a periodic 1D rhombic lattice, where each lattice site consists of two internal spin modes, by utilizing the combined synthetic dimensions of multiple internal levels and the motional state of a \(^{40}\mathrm{Ca}^{+}\) ion. We synthesize the non-Abelian U(2) gauge field by controlling the amplitude and the phase of the laser fields. From numerical simulation, we observe that the localization occurs when the interference matrix is nilpotent, or when the initial state is specifically set, which is absent in the Abelian case. The calculation parameters are set within the experimentally accessible range. Our simulation shows that the non-Abelian AB caging can be clearly observed in the time domain through the qudit-phonon joined detection. Our experimental scheme demonstrates the ability of trapped ions for the study of the non-Abelian physics \cite{chiral, dong24, synthesis2019, breach24, wang12, patrick23, wu2019non, shan24, neef23, zhang14, you2022observation, cpl-nonabelian1, cpl-nonabelian2} and showcases its potential for further research on exotic topological phenomena \cite{solid1,solid2,3dyudapeng, guangbodao1,guangbodao2, yanbo1, cpl-topo1, cpl-topo2, cpl-topo3}.

\textit{Theory of Non-Abelian AB Caging.} We consider a periodic one-dimensional (1D) rhombic lattice sketched in Fig.~\ref{fig:1}(a) with each site consisting of two (pseudo)spin modes \cite{original2,PRA}. The system in the presence of a U(2) background gauge field is described by the following Hamiltonian:
\begin{equation} \label{eq:1}
    \begin{split}
        H &= J\sum_{n = 0} \big(
            \bm{b}^{\dagger}_{\bm{n}}\bm{U}_{\bm{1}}\bm{a}_{\bm{n}} + 
            \bm{a}^{\dagger}_{\bm{n + 1}}\bm{U}_{\bm{2}}\bm{b}_{\bm{n}} 
        \\
        &\phantom{{}J\sum_{n = 0} (}+\bm{c}^{\dagger}_{\bm{n}}\bm{U}_{\bm{3}}\bm{a}_{\bm{n}} + 
            \bm{a}^{\dagger}_{\bm{n + 1}}\bm{U}_{\bm{4}}\bm{c}_{\bm{n}} + \text{h.c.}
        \big),
    \end{split}
\end{equation}
 \noindent where \(J\) is the uniform hopping strength, \(\bm{a_n}=[a_{\uparrow n}, a_{\downarrow n}]^{\mathrm{T}}\) (\(\bm{b_n}=[b_{\uparrow n}, b_{\downarrow n}]^{\mathrm{T}}\), \(\bm{c_n}=[c_{\uparrow n}, c_{\downarrow n}]^{\mathrm{T}}\)) with \(a_{\uparrow/\downarrow n}\) (\(b_{\uparrow/\downarrow n}\), \(c_{\uparrow/\downarrow n}\)) the annihilation operator for a particle at the \(\uparrow/\downarrow\) spin of the A (B, C) site of the \(n\)th unit cell (denoted as \(A_{\uparrow/\downarrow n}\) (\(B_{\uparrow/\downarrow n}\), \(C_{\uparrow/\downarrow n}\)) in the following), and \(\bm{U_i} \in \mathrm{U}(2)\) (\(i\)=1, 2, 3, and 4) is the link variable describing the unitary transformation experienced by a particle hopping between linked sites shown in Fig.~\ref{fig:1}(a). In the presence of such a gauge field, a charged particle is enabled to hop between lattice sites. As time progresses, its wave function typically spreads along the entire lattice chain. However, our interest lies in the situation where the wave function is confined to a finite region due to destructive interference and the particle cannot access sites outside this restricted zone, which is known as the AB caging effect. Unlike the Abelian U(1) situation, where the AB caging effect occurs when \(\pi\) magnetic flux is penetrated in each loop of the lattice \cite{original2}, here both \(\bm{U}\) and \(\bm{a_n}\) are matrix-valued, the caging condition needs to be generalized. With the definition of the rightward-moving interference matrix \cite{PRA}:
\begin{equation} \label{eq:2}
    \bm{I}=\frac{1}{2}(\bm{U_2U_1}+\bm{U_4U_3}),
\end{equation}
the AB caging effect occurs under the condition that \(\bm{I}\) is nilpotent, i.e.,
\begin{equation} \label{eq:3}
    \exists\ m\in\mathbb{N^*},\ \text{such that}\ \bm{I}^{m-1}=0,\ \bm{I}^{m}\neq0,
\end{equation}
and the nilpotency index of the interference matrix \(\bm{I}\) (denoted as \(\text{index}(\bm{I})\) in the following) is \(m\). For a particle initially populated in site \(\mathrm{A}_n\) moving rightward to \(\mathrm{A}_{n+1}\), it undergoes unitary transformation \(\bm{U_2U_1}\) along the up path and \(\bm{U_4U_3}\) along the down path, as shown in Fig.~\ref{fig:1}(a).
\(\bm{I}\) describes the interference of these two paths. Similarly, \(\bm{I}^{\dagger}=\frac{1}{2}(\bm{U_1^{\dagger}U_2^{\dagger}}+\bm{U_3^{\dagger}U_4^{\dagger}})\) describes the interference of the particle moving leftward. However, since \(\text{index}(\bm{I})=\text{index}(\bm{I}^{\dagger})\), we only need to focus on one of them in the discussion.
To illustrate the role that \(\text{index}(\bm{I})\) plays and for the convenience of the subsequent discussion, we here define the caging size \(s\) of the AB caging effect as follows. For a particle initially populated in site \(\mathrm{A}_n\). If its wave function can spread rightward to site \(\mathrm{A}_{n+s_{\mathrm{r}}-1}\) and cannot spread to site \(\mathrm{A}_{n+s_{\mathrm{r}}}\), then the rightward caging size is \(s_{\mathrm{r}}\). If its wave function can spread to site \(\mathrm{A}_{n-s_{\mathrm{l}}+1}\) and cannot spread leftward to site \(\mathrm{A}_{n-s_{\mathrm{l}}}\), then the leftward caging size is \(s_{\mathrm{l}}\). The caging size \(s\) is defined as \(s=\max\{s_{\mathrm{r}},s_{\mathrm{l}}\}\). It is evident that the caging size is limited by the nilpotency index of the interference matrix \(\bm{I}\), i.e., \(s\leq \text{index}(\bm{I})\).

In addition, the non-Abelian AB caging effect demands that the applied U(2) gauge field is non-Abelian in nature.
A criterion for distinguishing between the Abelian and the non-Abelian cases is proposed by introducing the loop operator \cite{goldman2009}:
\begin{equation} \label{eq:4}
    \bm{W}=\bm{U_3}^{\dagger}\bm{U_4}^{\dagger}\bm{U_2}\bm{U_1},
\end{equation}
which is the unitary transformation for particles hopping around a cell of the lattice starting from site A in clockwise direction. If \(\bm{W}\propto\hat{\bm{1}}\), where \(\hat{\bm{1}}\) is the identity matrix, then the AB caging effect is in the Abelian regime. In this case, particles starting from different spin modes, \(\mathrm{A}_{\uparrow}\) and \(\mathrm{A}_{\downarrow}\) respectively, would accumulate the same AB phase, and our problem decouples into two identical Abelian AB caging problems. On the other hand, if \(\bm{W} \not\propto \hat{\bm{1}}\), then the AB caging effect is in the non-Abelian regime.

Based on this criterion, we can deduce three features which are exclusive to non-Abelian AB caging effects. 
Firstly, when the interference matrix \(\bm{I}\) is nilpotent, the nilpotency index can be greater than one. As a result, the caging size can be larger than one.
Secondly, the rightward caging size and the leftward caging size can be different, and this leftward-rightward asymmetry can be altered by the initial state.
Lastly, even when the interference matrix \(\bm{I}\) is not nilpotent, the AB caging effect can still occur with certain specific initial states.
These features will not occur in Abelian cases (see Appendix for details), and therefore the observation of these features is the most direct manifestation of the non-Abelian AB caging effect.

\begin{figure*}[t]
    \centering
    \begin{subfigure}{0.8\textwidth}
        \begin{overpic}[width=1\textwidth]{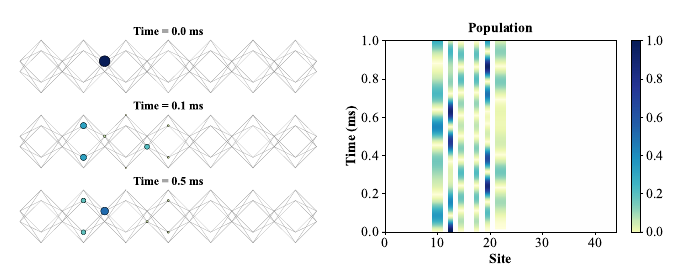}
            \put(0,37){(a)}
        \end{overpic}
        \label{fig:2_subfig_a}
    \end{subfigure}
    \begin{subfigure}{0.8\textwidth}
        \begin{overpic}[width=1\textwidth]{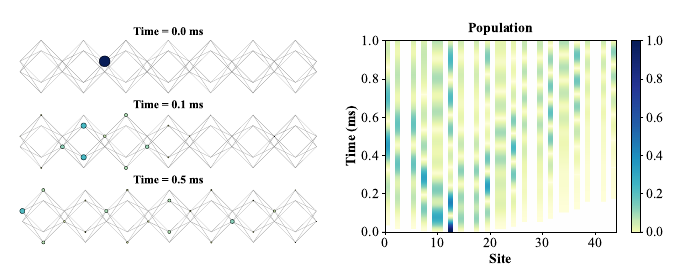}
            \put(0,37){(b)}
        \end{overpic}
        \label{fig:2_subfig_b}
    \end{subfigure}
    \caption{Simulation of the non-Abelian AB caging effect by laser-ion interaction, with initial state being \(\mathrm{A_{\uparrow 2}}\) and different link variables. Parameters are \(\omega=2\pi\)×2 MHz, \(\eta = 0.1\), \(J/h=2.5\ \text{kHz}\), as defined in the text. (a) \(\bm{U_1}=\bm{U_4}=\tiny\begin{pmatrix}1 & 0\\0 & 1\end{pmatrix}\), \(\bm{U_2}=\tiny\begin{pmatrix}0 & 1\\1 & 0\end{pmatrix}\), \(\bm{U_3}=\tiny\begin{pmatrix}0 & 1\\-1 & 0\end{pmatrix}\), where \(\text{index}(\bm{I})=2\). (b) \(\bm{U_1}=\bm{U_4}=\tiny\begin{pmatrix}1 & 0\\0 & 1\end{pmatrix}\), \(\bm{U_2}=\tiny\begin{pmatrix}0 & \mathrm{i}\\\mathrm{i} & 0\end{pmatrix}\), and \(\bm{U_3}=\tiny\begin{pmatrix}0 & 1\\-1 & 0\end{pmatrix}\), where \(\bm{I}\) is non-nilpotent. For each subfigure, the left panels show probability distribution of each site (indicated by the area of the circle as well as the filled color) at \(t=0\) ms, \(t=0.1\) ms and \(t=0.5\) ms, and the right panels show probability evolution of each site from 0 ms to 1 ms with sites numbered in the order of: \(\mathrm{A_{\uparrow 0}, A_{\downarrow 0}, B_{\uparrow 0}, B_{\downarrow 0}, C_{\uparrow 0}, C_{\downarrow 0}, A_{\uparrow 1}},..., C_{\downarrow 6}, A_{\uparrow 7},A_{\downarrow 7}. \) The color bar indicates the value of the probability.}
    \label{fig:2}
\end{figure*}

\textit{Experimental Scheme for a Trapped Ion.} We consider a single trapped \(^{40}\mathrm{Ca}^{+}\) ion in a linear Paul trap. 
In our experimental scheme, we utilize the synthetic dimensions of ionic internal states including two Zeeman sub-levels of the ground state $\ket{S_{1/2},m_{j}=-1/2,1/2}$ (denoted as $\ket{g=1,2}$ in the following) and four Zeeman sub-levels of the metastable state $\ket{D_{5/2},m_{j}=-3/2,-1/2,1/2,3/2}$ (denoted as $\ket{e=3,4,5,6}$ in the following), as well as motional Fock states $\ket{n}$ of the ion, as shown in Fig.~\ref{fig:1}(b). In one unit cell of the lattice, we encode site A utilizing \(\ket{g=1,2}\) internal states, site B utilizing \(\ket{e=3,4}\) internal states and site C utilizing \(\ket{e=5,6}\) internal states. The replication of this unit cell along the lattice axis is achieved through different phonon Fock states $\ket{n=0,1,2,...}$ of the ions.
The encoding protocol between \(\{\ket{g}\otimes\ket{n}, \ket{e}\otimes\ket{n}\}\) and \(\{\mathrm{A}_{\uparrow(\downarrow) n}, \mathrm{B}_{\uparrow\downarrow n},\mathrm{C}_{\uparrow(\downarrow) n}\}\) is shown in Fig.~\ref{fig:1}(c).
As for the measurement of the probability distribution of each site, we can utilize the electron shelving technique \cite{ringbauer2022} and blue sideband transition \cite{meekhof1996generation} to accomplish the readout of the 6-level qudit and the phonon state (see Appendix for details).

Hopping terms between sites can be achieved through laser-ion interaction coupling the multiple internal levels of the ion and the harmonic oscillator model of its motion. In the Lamb-Dicke limit where the motion of the ion is negligible compared with the laser wavelength, we can tune the lasers to coherently drive $\ket{g}\otimes\ket{n}\leftrightarrow\ket{e}\otimes\ket{n}$ namely carrier (CAR) transition, and $\ket{g}\otimes\ket{n+1}\leftrightarrow\ket{e}\otimes\ket{n}$ namely red sideband (RSB) transition. In accordance with the coding protocol in Fig.~\ref{fig:1}(c), we utilize eight of each transition and create the Hamiltonian in the ion's rest frame \cite{bible}: 
\begin{widetext}
\begin{equation}
H_{\text{I}}=\sum_{e=3}^{6}\sum_{g=1,2}\hbar\varOmega_{g,e}^{\text{car}}\big(\ket{e}\bra{g}\mathrm{e}^{\mathrm{i}\phi_{g,e}^{\text{car}}}\otimes {I}_{\mathrm{ph}}+\text{h.c.})+ \sum_{e=3}^{6}\sum_{g=1,2}\hbar\eta\varOmega_{g,e}^{\text{rsb}}(\mathrm{i}\ket{e}\bra{g}\mathrm{e}^{\mathrm{i}\phi_{g,e}^{\text{rsb}}}\otimes {a}_{\mathrm{ph}}+\text{h.c.}\big),
 \label{eq:5}
\end{equation}
\end{widetext}
where \(\hbar\) is the reduced Planck constant, \(\varOmega_{g,e}^{\text{car}}\) (\(\varOmega_{g,e}^{\text{rsb}}\)) and \(\phi_{g,e}^{\text{car}}\) (\(\phi_{g,e}^{\text{rsb}}\)) are the coupling strength and phase of CAR (RSB) transitions linking internal state $\ket{g}$ and $\ket{e}$, \(\eta\) is the Lamb-Dicke factor, \({I}_{\mathrm{ph}}\) is the phonon identity operator, and \({a}_{\mathrm{ph}}\) is the phonon lowering operator.
By setting 

\begin{subequations}
\label{eq:6}
\begin{equation}
\frac{\hbar}{J}\begin{pmatrix} 
\varOmega_{1,3}^{\text{car}}\mathrm{e}^{\mathrm{i}\phi_{1,3}^{\text{car}}}& 
\varOmega_{1,4}^{\text{car}}\mathrm{e}^{\mathrm{i}\phi_{1,4}^{\text{car}}}\\
\varOmega_{2,3}^{\text{car}}\mathrm{e}^{\mathrm{i}\phi_{2,3}^{\text{car}}} & 
\varOmega_{2,4}^{\text{car}}\mathrm{e}^{\mathrm{i}\phi_{2,4}^{\text{car}}}
\end{pmatrix}=\bm{U_1},
\end{equation}
\begin{equation}
\frac{\hbar}{J}\begin{pmatrix} 
\varOmega_{1,5}^{\text{car}}\mathrm{e}^{\mathrm{i}\phi_{1,5}^{\text{car}}}& 
\varOmega_{1,6}^{\text{car}}\mathrm{e}^{\mathrm{i}\phi_{1,6}^{\text{car}}}\\
\varOmega_{2,5}^{\text{car}}\mathrm{e}^{\mathrm{i}\phi_{2,5}^{\text{car}}} & 
\varOmega_{2,6}^{\text{car}}\mathrm{e}^{\mathrm{i}\phi_{2,6}^{\text{car}}}
\end{pmatrix}=\bm{U_3},
\end{equation}
\begin{equation}
\mathrm{i}\eta\frac{\hbar}{J}\begin{pmatrix} 
\varOmega_{1,3}^{\text{rsb}}\mathrm{e}^{\mathrm{i}\phi_{1,3}^{\text{rsb}}}& 
\varOmega_{2,3}^{\text{rsb}}\mathrm{e}^{\mathrm{i}\phi_{2,3}^{\text{rsb}}}\\
\varOmega_{1,4}^{\text{rsb}}\mathrm{e}^{\mathrm{i}\phi_{1,4}^{\text{rsb}}} & 
\varOmega_{2,4}^{\text{rsb}}\mathrm{e}^{\mathrm{i}\phi_{2,4}^{\text{rsb}}}
\end{pmatrix}=\bm{U_2},
\end{equation}
\begin{equation}
\mathrm{i}\eta\frac{\hbar}{J}\begin{pmatrix} 
\varOmega_{1,5}^{\text{rsb}}\mathrm{e}^{\mathrm{i}\phi_{1,5}^{\text{rsb}}}& 
\varOmega_{2,5}^{\text{rsb}}\mathrm{e}^{\mathrm{i}\phi_{2,5}^{\text{rsb}}}\\
\varOmega_{1,6}^{\text{rsb}}\mathrm{e}^{\mathrm{i}\phi_{1,6}^{\text{rsb}}} & 
\varOmega_{2,6}^{\text{rsb}}\mathrm{e}^{\mathrm{i}\phi_{2,6}^{\text{rsb}}}
\end{pmatrix}=\bm{U_4},
\end{equation}
\end{subequations}

\noindent Eq.~(\ref{eq:5}) becomes:

\begin{equation} \label{eq:7}
    \begin{split}
        H_{\text{I}} &= J\sum_{n = 0} \big(
            \bm{b}^{\dagger}_{\bm{n}}\bm{U}_{\bm{1}}\bm{a}_{\bm{n}} + 
            \sqrt{n}\bm{a}^{\dagger}_{\bm{n + 1}}\bm{U}_{\bm{2}}\bm{b}_{\bm{n}} 
        \\
        &\phantom{{}J\sum_{n = 0}\sqrt{n} (}+\bm{c}^{\dagger}_{\bm{n}}\bm{U}_{\bm{3}}\bm{a}_{\bm{n}} + 
            \sqrt{n}\bm{a}^{\dagger}_{\bm{n + 1}}\bm{U}_{\bm{4}}\bm{c}_{\bm{n}} + \text{h.c.}
        \big).
    \end{split}
\end{equation}

We can see a realization of Eq.~(\ref{eq:1}) except for the \(\sqrt{n}\) amplitude modulation. This modulation originates from the intrinsic property of the phonon, as we partly use a boson to simulate a fermion. However, it does not affect essential signatures of non-Abelian AB caging, while only accelerating the evolution as \(n\) increases. 
Since \(\varOmega_{g,e}^{\text{car}}\) (\(\varOmega_{g,e}^{\text{rsb}}\)) and \(\phi_{g,e}^{\text{car}}\) (\(\phi_{g,e}^{\text{rsb}}\)) can be set by changing the laser's amplitude and phase via acoustic-optical modulators, arbitrary \(\bm{U_1, U_2, U_3}\) and \(\bm{U_4}\) matrices can be realized. Up to this point, we have successfully constructed a synthetic U(2) gauge field on a single ion. 
This setup is now primed for us to delve into the non-Abelian AB caging effect.

\begin{figure*}[t]
    \centering
    \begin{subfigure}{0.8\textwidth}
        \begin{overpic}[width=1\textwidth]{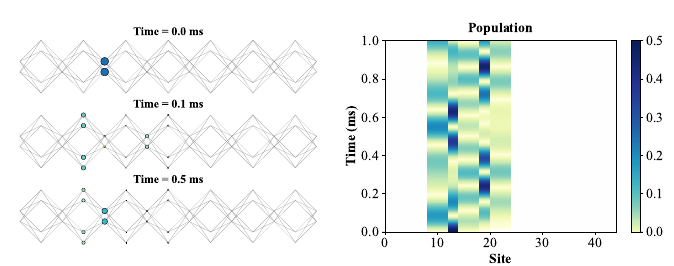}
            \put(0,37){(a)}
        \end{overpic}
        \label{fig:2_subfig_a}
    \end{subfigure}
    \begin{subfigure}{0.8\textwidth}
        \begin{overpic}[width=1\textwidth]{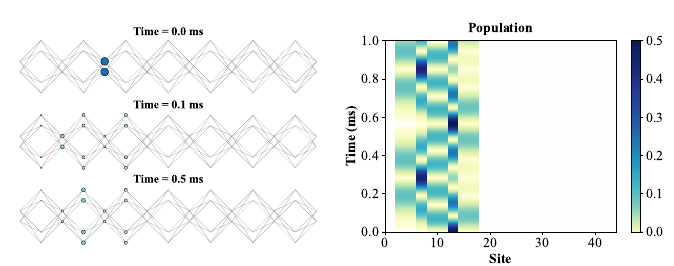}
            \put(0,37){(b)}
        \end{overpic}
        \label{fig:2_subfig_b}
    \end{subfigure}
    
    \caption{Simulation of the non-Abelian AB caging effect by laser-ion interaction, with \(\bm{U_1}=\bm{U_4}=\tiny\begin{pmatrix}1 & 0\\0 & 1\end{pmatrix}\), \(\bm{U_2}=\tiny\begin{pmatrix}0 & \mathrm{i}\\\mathrm{i} & 0\end{pmatrix}\), \(\bm{U_3}=\tiny\begin{pmatrix}1 & 0\\0 & -1\end{pmatrix}\) where \(\text{index}(\bm{I})=2\), and different initial states. Parameters and figure formats are the same as those in Fig.~\ref {fig:2}. (a) Initial state is \(\frac{\mathrm{A_{\downarrow 2}}+\mathrm{i}\mathrm{A_{\uparrow 2}}}{\sqrt{2}}\). (b) Initial state is \(\frac{\mathrm{A_{\downarrow 2}}-\mathrm{i}\mathrm{A_{\uparrow 2}}}{\sqrt{2}}\).}
    \label{fig:3}
\end{figure*}

\textit{Numerical Simulation of Non-Abelian AB Caging Effect.} As mentioned above, we need to first verify the feasibility of this scheme under experimental parameters. For a single trapped \(^{40}\mathrm{Ca}^{+}\) ion in a linear Paul trap, the motional trap frequency is usually around \(\omega=2\pi\)×2 MHz, the Lamb-Dicke parameter \(\eta\) of \(^{40}\mathrm{Ca}^{+}\) ion driven by the 729 nm laser is set to 0.1 as an example. We set \(J/h=2.5\ \text{kHz}\), via calibrating \(\pi\)-time of CAR and RSB transitions in the first cell to 0.2 ms, which is consistent with the usual \(\pi\)-time in ion trap experiments \cite{others}. The most severe unwanted off-resonant excitation comes from RSB transitions driving their corresponding carrier transitions, with the frequency detuning $\omega$.
Under the chosen coupling strength, the population excited due to off-resonant excitation can be estimated as \(\frac{(\varOmega_{1,3}^{\text{rsb}})^2}{(\varOmega_{1,3}^{\text{rsb}})^2+{\omega}^2} \sim 1.6×10^{-4} \ll1\), and thus can be neglected \cite{foot2005atomic}. We analyze the non-Abelian AB caging behavior in the time domain by calculating the probability distribution of the lattice sites varying with time, corresponding to the internal-motional state joint detection in ion trap experiments \cite{ringbauer2022,meekhof1996generation} (see Appendix for details). Considering the accuracy of phonon detection and without affecting the observation of phenomena, we truncate the phonon detection up to \(n=7\).

In this work, we observe three features of non-Abelian AB caging discussed in the previous context through three sets of simulations.
Firstly, we demonstrate the non-Abelian AB caging effect with the caging size \(s=2\). In Fig.~\ref{fig:2}(a), we apply \(\bm{U_1}=\bm{U_4}=
\tiny\begin{pmatrix}1 & 0\\0 & 1\end{pmatrix}\), \(\bm{U_2}=\tiny\begin{pmatrix}0 & 1\\1 & 0\end{pmatrix}\), \(\bm{U_3}=\tiny\begin{pmatrix}0 & 1\\-1 & 0\end{pmatrix}\), where \(\text{index}(\bm{I})=2\). The particle starts from \(\mathrm{\mathrm{A_{\uparrow 2}}}\) and the wave function is confined to the nearest and next-nearest cells adjacent to \(\mathrm{A_{\uparrow 2}}\), suggesting \(s=2\). By contrast, in Fig.~\ref{fig:2}(b), we apply \(\bm{U_1}=\bm{U_4}=
\tiny\begin{pmatrix}1 & 0\\0 & 1\end{pmatrix}\), \(\bm{U_2}=\tiny\begin{pmatrix}0 & \mathrm{i}\\\mathrm{i} & 0\end{pmatrix}\), and \(\bm{U_3}=\tiny\begin{pmatrix}0 & 1\\-1 & 0\end{pmatrix}\), where \(\bm{I}\) is non-nilpotent. In this case, the wave function is no longer confined to a finite region and spreads along the entire lattice.

\begin{figure*}[t]
    \centering
    \begin{subfigure}{0.8\textwidth}
        \begin{overpic}[width=1\textwidth]{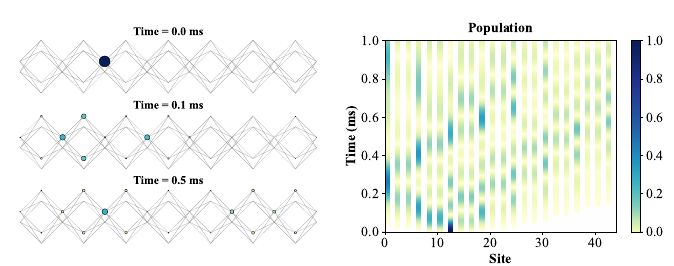}
            \put(0,37){(a)}
        \end{overpic}
        \label{fig:2_subfig_a}
    \end{subfigure}
    \begin{subfigure}{0.8\textwidth}
        \begin{overpic}[width=1\textwidth]{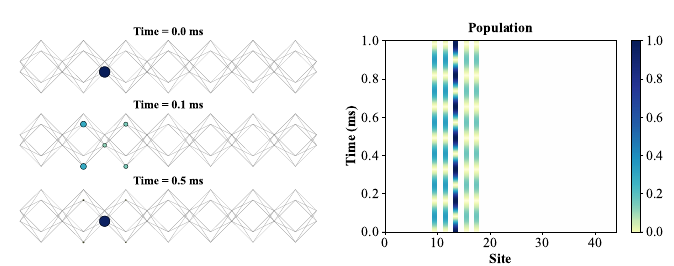}
            \put(0,37){(b)}
        \end{overpic}
        \label{fig:2_subfig_b}
    \end{subfigure}
    \caption{Simulation of the non-Abelian AB caging effect by laser-ion interaction, with \(\bm{U_1}=\bm{U_2}=\bm{U_4}=\tiny\begin{pmatrix}1 & 0\\0 & 1\end{pmatrix}\), \(\bm{U_3}=\tiny\begin{pmatrix}1 & 0\\0 & -1\end{pmatrix}\) where \(\bm{I}\) is non-nilpotent, and different initial states. Parameters and figure formats are the same as those in Fig.~\ref {fig:2}. (a) Initial state is \(\mathrm{A_{\uparrow 2}}\). (b) Initial state is \(\mathrm{A_{\downarrow 2}}\).}
    \label{fig:4}
\end{figure*}

Secondly, we focus on the leftward-rightward asymmetry feature and its sensitivity to the initial state. We set \(\bm{U_1} = \bm{U_4}=\tiny\begin{pmatrix}1 & 0\\0 & 1\end{pmatrix}\), \(\bm{U_2}=\tiny\begin{pmatrix}0 & \mathrm{i}\\\mathrm{i} & 0\end{pmatrix}\), and \(\bm{U_3}=\tiny\begin{pmatrix}1 & 0\\0 & -1\end{pmatrix}\), where \(\text{index}(\bm{I})=2\). When the initial state is \(\frac{\mathrm{A_{\downarrow 2}}+\mathrm{i}\mathrm{A_{\uparrow 2}}}{\sqrt{2}}\), the wave function is confined to the region with two rightward cells and one leftward cell away from the initial site, as shown in Fig.~\ref{fig:3}(a), indicating \(s_{\mathrm{r}}=2\) and \(s_{\mathrm{l}}=1\). In comparison, when the initial state is \(\frac{\mathrm{A_{\downarrow 2}}-\mathrm{i}\mathrm{A_{\uparrow 2}}}{\sqrt{2}}\), the wave function is confined to the region with one rightward cell and two leftward cells away from the initial site, as shown in Fig.~\ref{fig:3}(b), indicating \(s_{\mathrm{r}}=1\) and \(s_{\mathrm{l}}=2\). These results confirm that the leftward-rightward asymmetry exists in non-Abelian AB caging and can be altered by the initial state. 

Finally, we observe the non-Abelian AB caging effect induced by a specific initial state in the case when the interference matrix \(\bm{I}\) is not nilpotent. Under \(\bm{U_1}=\bm{U_2}=\bm{U_4}=\tiny\begin{pmatrix}1 & 0\\0 & 1\end{pmatrix}\), \(\bm{U_3}=\tiny\begin{pmatrix}1 & 0\\0 & -1\end{pmatrix}\), where \(\bm{I}\) is non-nilpotent, initial state being \(\mathrm{A_{\uparrow 2}}\) exhibits no caging effect as shown in Fig.~\ref{fig:4}(a). However, changing the initial state to \(\mathrm{A_{\uparrow 2}}\) makes the caging effect revive with the caging size \(s=1\), as shown in Fig.~\ref{fig:4}(b). These results show that even when the interference matrix \(\bm{I}\) is not nilpotent, the non-Abelian AB caging effect can still occur with certain choices of initial states. It is noted that all the link variables of the gauge fields applied in our study satisfy \(\bm{W}=\bm{U_3}^{\dagger}\bm{U_4}^{\dagger}\bm{U_2}\bm{U_1}\not\propto \hat{\bm{1}}\), guaranteeing the non-Abelian nature of the AB caging effects.

\begin{figure*}[t]
    \centering
    \begin{subfigure}{0.4\textwidth}
        \begin{overpic}[width=\textwidth]{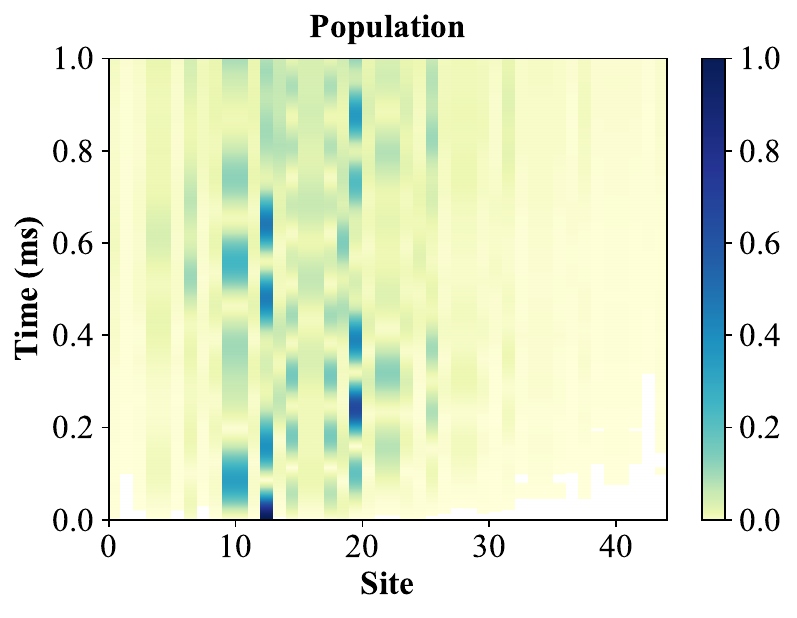}
            \put(0,76){(a)}
        \end{overpic}
        \label{fig:5_subfig_a}
    \end{subfigure}
    \begin{subfigure}{0.4\textwidth}
        \begin{overpic}[width=\textwidth]{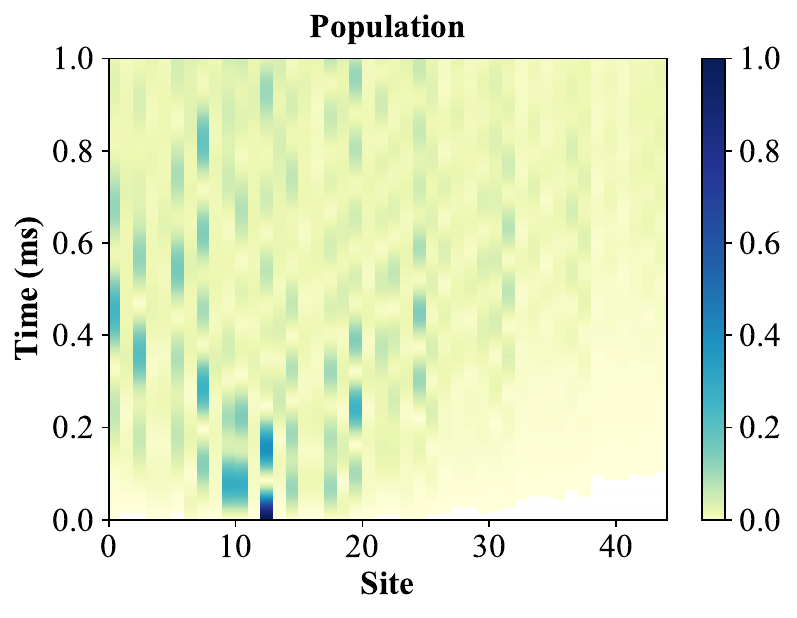}
            \put(0,76){(b)}
        \end{overpic}
        \label{fig:5_subfig_b}
    \end{subfigure}
    \caption{Simulation of the non-Abelian AB caging effect by laser-ion interaction under the full Lindblad master equation, with initial state being \(\mathrm{A_{\uparrow 2}}\) and different link variables. Parameters are \(\omega = 2\pi\)×2 MHz, \(\eta = 0.1\), \(J/h = 2.5\ \text{kHz}\), \(\dot{\overline{n}}=0.2\) quantas/s, \(T_2^{\text{m}}=35\) ms, and \(T_2^{\text{s}}=40\) ms, as defined in the text. Figure formats are the same as those in Fig.~\ref {fig:2}. (a) \(\bm{U_1}=\bm{U_4}=\tiny\begin{pmatrix}1 & 0\\0 & 1\end{pmatrix}\), \(\bm{U_2}=\tiny\begin{pmatrix}0 & 1\\1 & 0\end{pmatrix}\), \(\bm{U_3}=\tiny\begin{pmatrix}0 & 1\\-1 & 0\end{pmatrix}\), where \(\text{index}(\bm{I})=2\). (b) \(\bm{U_1}=\bm{U_4}=\tiny\begin{pmatrix}1 & 0\\0 & 1\end{pmatrix}\), \(\bm{U_2}=\tiny\begin{pmatrix}0 & \mathrm{i}\\\mathrm{i} & 0\end{pmatrix}\), and \(\bm{U_3}=\tiny\begin{pmatrix}0 & 1\\-1 & 0\end{pmatrix}\), where \(\bm{I}\) is non-nilpotent.}
    \label{fig:5}
\end{figure*}

\textit{Effects with Experimental Imperfections.} In the above numerical simulation, we disregard the decoherence of the oscillator mode caused by motional heating and dephasing, as well as the decoherence of the spin state. These phenomena represent the primary sources of noise in ion trap systems. To evaluate the influence of such noise factors in experimental setups, we take into account the complete Lindblad master equation:
\begin{equation} \label{eq:8}
    \frac{d\rho}{dt}=\frac{1}{\mathrm{i}\hbar}[H_{\text{I}},\rho]+\sum_{j}(L_j \rho L_j^{\dagger}-\frac{1}{2}
    L_j^{\dagger}L_j\rho-\frac{1}{2} \rho L_j^{\dagger}L_j),
\end{equation}
in which \(\rho\) is the density matrix of the system, the dissipation operators are $L_{\text{heat}}=\sqrt{\dot{\overline{n}}} a_{\mathrm{ph}}^{\dagger}$, $L_{\text{cool}}=\sqrt{\dot{\overline{n}}} a_{\mathrm{ph}}$, $L_{\text{motion}}=\sqrt{1/T_2^{\text{m}}} a_{\mathrm{ph}}^{\dagger}a_{\mathrm{ph}}$, and $L_{\text{spin}}=\sqrt{1/T_2^{\text{s}}} \sum_{e}\sum_{g}(\ket{e}\bra{e}-\ket{g}\bra{g})$. Here $\dot{\overline{n}}$ represents the heating rate of the oscillator, $T_2^{\text{m}}$ represents the motional dephasing time, and $T_2^{\text{s}}$ represents the spin qudit dephasing time. For the following numerical simulation, we incorporate these imperfections with some values of $\dot{\overline{n}}=0.2\ \text{quantas}/s$, $T_2^{\text{m}}=35\ \text{ms}$ \cite{T2m}, and $T_2^{\text{s}}=40\ \text{ms}$ \cite{T2s}, which are typical for ion trap platforms.

Including these possible decoherence sources, we show numerical simulation results on the non-Abelian AB caging effect under the full Lindblad master equation. With the initial state set to \(\mathrm{A_{\uparrow 2}}\), in Fig.~\ref{fig:5}(a), we apply \(\bm{U_1}=\bm{U_4}=\tiny\begin{pmatrix}1 & 0\\0 & 1\end{pmatrix}\), \(\bm{U_2}=\tiny\begin{pmatrix}1 & 0\\0 & -1\end{pmatrix}\), \(\bm{U_3}=\tiny\begin{pmatrix}0 & -1\\1 & 0\end{pmatrix}\), where \(\text{index}(\bm{I})=2\), which is consistent with Fig.~\ref{fig:2}(a), so that the caging size should be two and the wave function should be confined to the nearest and next-nearest cells adjacent to \(\mathrm{A_{\uparrow 2}}\). It can be observed that with the imperfections in the experiment, there appears a slight leakage of the probability distribution outside this region, but a caging pattern of size two is still observable. 
In Fig.~\ref{fig:5}(b), we apply \(\bm{U_1}=\bm{U_4}=
\tiny\begin{pmatrix}1 & 0\\0 & 1\end{pmatrix}\), \(\bm{U_2}=\tiny\begin{pmatrix}0 & \mathrm{i}\\\mathrm{i} & 0\end{pmatrix}\), and \(\bm{U_3}=\tiny\begin{pmatrix}0 & 1\\-1 & 0\end{pmatrix}\), where \(\bm{I}\) is non-nilpotent. We observe no non-Abelian AB caging effect and the result is similar to Fig.~\ref{fig:2}(b).
Compared with Fig.~\ref{fig:5}(b), Fig.~\ref{fig:5}(a) still provides recognizable non-Abelian AB caging patterns in addition to slight probability distribution leakage outside the theoretical caging sites, which we attribute to the alternation of the dynamics due to decoherence.
These numerical simulation results confirm that even with experimental imperfections, the signatures of the non-Abelian AB caging effect can still be observed and manipulated through non-Abelian links between sites and initial states, showing the feasibility of our scheme.

In conclusion, we have proposed an experimental scheme to realize non-Abelian AB caging on the 1D rhombic lattice created by internal and motional states of a single trapped ion. With numerical simulations, we demonstrate the non-Abelian AB caging effect with novel features exclusive to the non-Abelian regime. These include the caging size of two, leftward-rightward asymmetry and caging effect induced by the initial state.
The realization of Fock-state lattices constructed by phonon number states of the motional mode of the ion shows an analogous method to that in superconducting qubits where photon number states are utilized \cite{3dyudapeng,cai2021topological,deng2022observing}.
The highly tunable experimental scheme provides new opportunities to the study of exotic phenomena related to gauge fields, including the topological states of matter \cite{breach24,goldman2014,chi32}, non-Abelian quantum holonomy \cite{neef23, shan24}, and the quantum Hall effect \cite{chi53, wang12}.
Besides, by considering multiple ions and motional modes, our work can be extended to higher dimensions and more complicated lattice structures, as well as higher dimensional non-Abelian U($N$) gauge fields.

\textit{Acknowledgments.} This work was supported by the National Natural Science Foundation of China (Grant No. 92165206, No. 12275090, No. 12304554) and Innovation Program for Quantum Science and Technology (Grant No. 2021ZD0301603, No. 2021ZD0302303).

\appendix

\section{Proof of non-Abelian AB caging features}

\begin{figure*}[t]
    \centering
    \begin{subfigure}{0.3\textwidth}
        \begin{overpic}[width=1\textwidth]{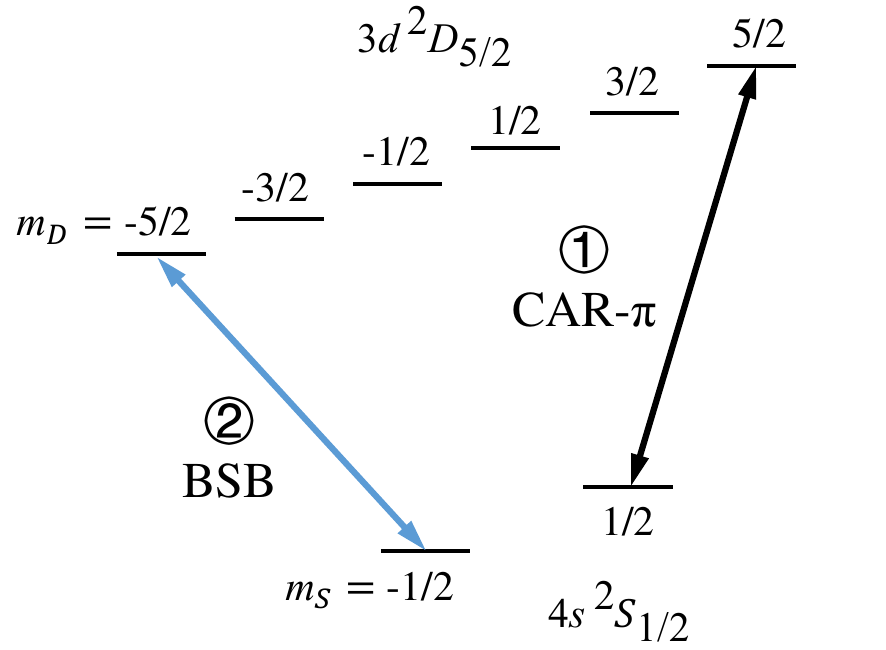}
            \put(0,65){(a)}
        \end{overpic}
    \end{subfigure}
    \begin{subfigure}{0.3\textwidth}
        \begin{overpic}[width=1\textwidth]{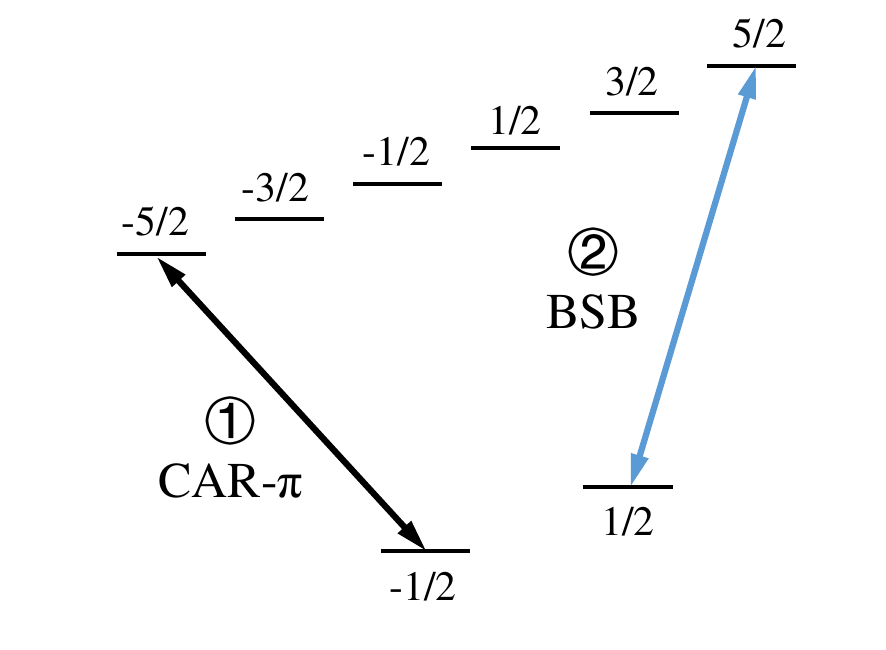}
            \put(0,65){(b)}
        \end{overpic}
    \end{subfigure}
    \begin{subfigure}{0.3\textwidth}
        \begin{overpic}[width=1\textwidth]{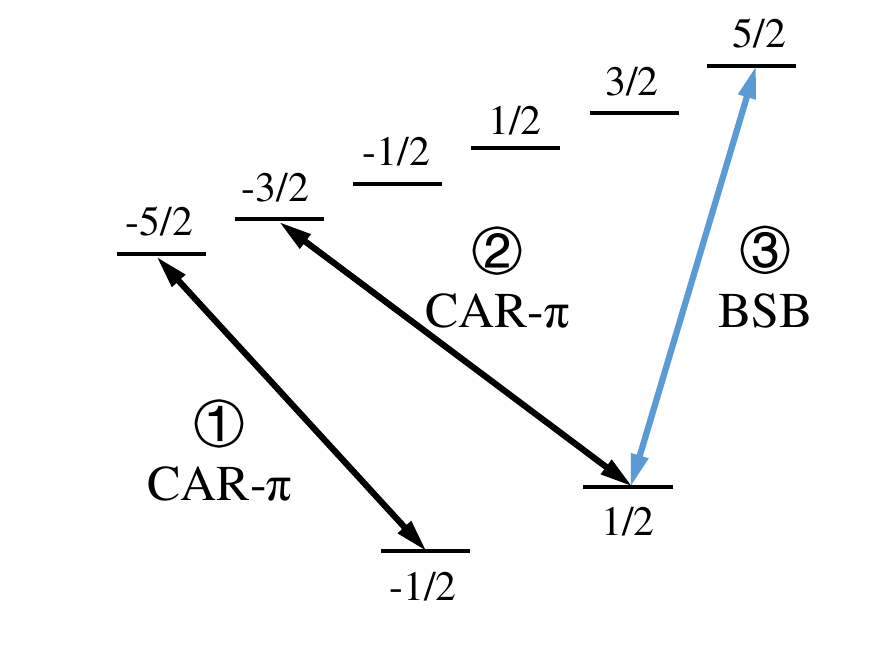}
            \put(0,65){(c)}
        \end{overpic}
    \end{subfigure}
    \begin{subfigure}{0.325\textwidth}
        \begin{overpic}[width=1\textwidth]{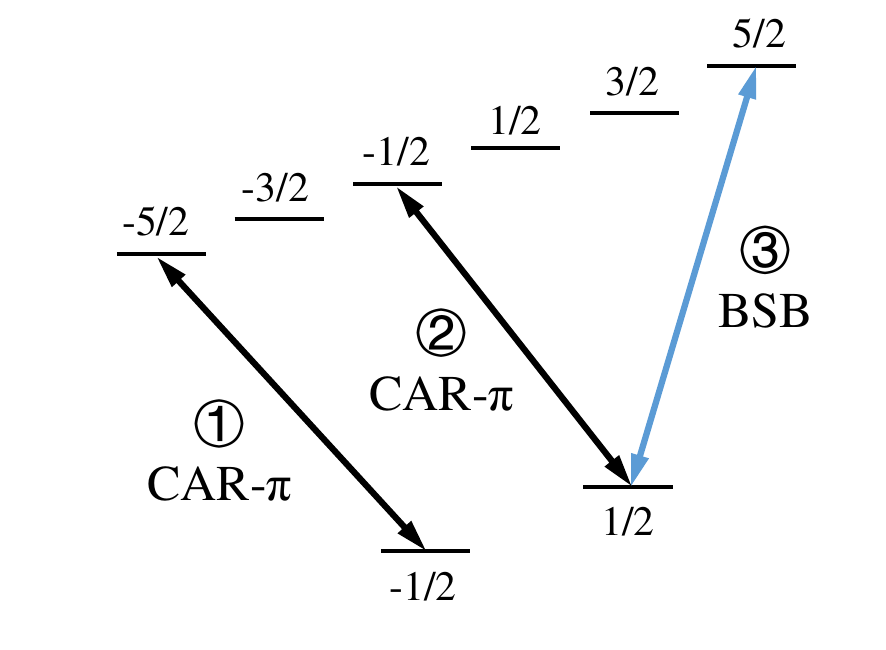}
            \put(0,65){(d)}
        \end{overpic}
    \end{subfigure}
    \begin{subfigure}{0.3\textwidth}
        \begin{overpic}[width=1\textwidth]{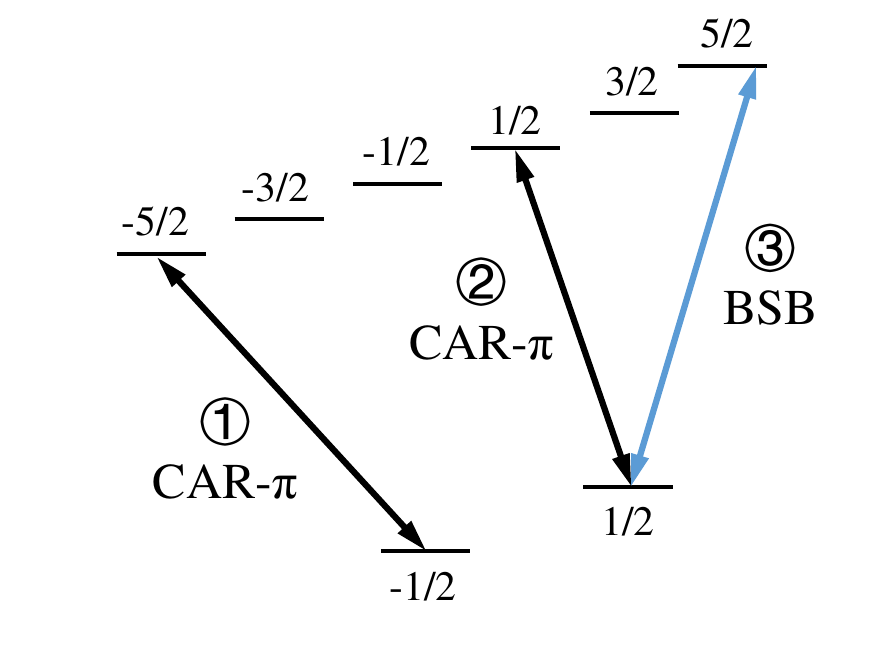}
            \put(0,65){(e)}
        \end{overpic}
    \end{subfigure}
    \begin{subfigure}{0.3\textwidth}
        \begin{overpic}[width=1\textwidth]{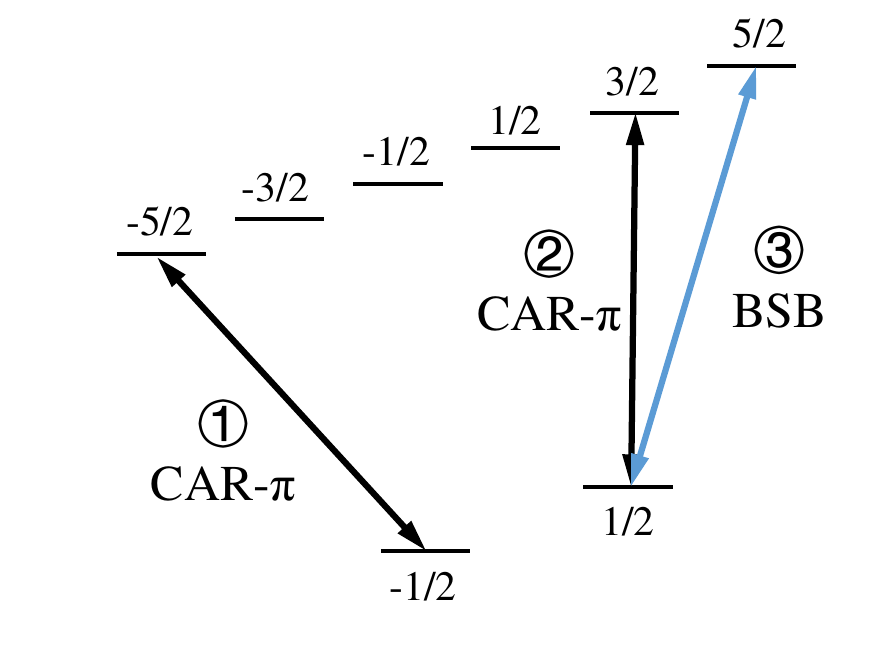}
            \put(0,65){(f)}
        \end{overpic}
    \end{subfigure}
    \caption{Electron shelving and blue sideband transition protocol for population measurement of (a) $\ket{S_{1/2},m_{j}=-1/2}\otimes\ket{n}$ state, (b) $\ket{S_{1/2},m_{j}=1/2}\otimes\ket{n}$ state, (c) $\ket{D_{5/2},m_{j}=-3/2}\otimes\ket{n}$ state, (d) $\ket{D_{5/2},m_{j}=-1/2}\otimes\ket{n}$ state, (e) $\ket{D_{5/2},m_{j}=1/2}\otimes\ket{n}$ state and (f) $\ket{D_{5/2},m_{j}=3/2}\otimes\ket{n}$ state. ``CAR-$\pi$" denotes resonant transition between internal states between the $S_{1/2}$ and $D_{5/2}$ manifolds. ``BSB" denotes blue sideband transition, which jointly excite the internal states and changing motional state by one quanta.}
    \label{fig:s}
\end{figure*}

In this Appendix, we prove the claim that the three AB caging features mentioned in the main text are exclusive to the non-Abelian case and do not exist in the Abelian case.

We prove the first feature by the method of contradiction. Assume that for the Abelian case, the interference matrix \(\bm{I}\) is nilpotent with the nilpotency index \(m>1\). According to the definition of the Abelian AB caging that \(\bm{W} \propto \hat{\bm{1}}\), without loss of generality, we set \(\bm{W}=\mathrm{e}^{\mathrm{i}\theta}\hat{\bm{1}}\), where \(\theta\) is an arbitrary angle. By considering \(\bm{W}=\bm{U_3}^{\dagger}\bm{U_4}^{\dagger}\bm{U_2}\bm{U_1}\), we can obtain:
\begin{equation} \label{eq:A1}
    \bm{U_2}\bm{U_1}=\mathrm{e}^{\mathrm{i}\theta}\bm{U_4}\bm{U_3},
\end{equation}
\begin{equation} \label{eq:A2}
    \bm{I}=\frac{1+\mathrm{e}^{\mathrm{i}\theta}}{2}\bm{U_4}\bm{U_3}.
\end{equation}
Since, according to the assumption that \(m>1\), we know that \(I\neq0\). Thus, \(\frac{1+\mathrm{e}^{\mathrm{i}\theta}}{2}\neq0\) and \(\big(\frac{1+\mathrm{e}^{\mathrm{i}\theta}}{2}\big)^m\neq0\). Also, by the properties of unitary matrices \(\bm{U_1, U_2, U_3}\) and \(\bm{U_4}\), \((\bm{U_4}\bm{U_3})^m\neq0\).
We calculate \(\bm{I}^m\):
\begin{equation} \label{eq:A3}
    \bm{I}^m=\big(\frac{1+\mathrm{e}^{\mathrm{i}\theta}}{2}\bm{U_4}\bm{U_3}\big)^m=\big(\frac{1+\mathrm{e}^{\mathrm{i}\theta}}{2}\big)^m(\bm{U_4}\bm{U_3})^m\neq0,
\end{equation}
which is in contradiction with the assumption that \(\bm{I}\) is a nilpotent matrix with nilpotency index \(m\). So our initial assumption is wrong, which completes the proof. 

The second feature is obvious. Since, as described in the first feature, the caging size of the Abelian AB cage can only be one, there is no leftward-rightward asymmetry behavior, not to mention the possibility of altering the pattern of leftward-rightward asymmetry via the initial state.

Finally, we prove the third feature also using the method of contradiction. Assume that for the Abelian case, the interference matrix \(\bm{I}\) is non-nilpotent, and there exists an integer \(m\) and an initial state \(\ket{\psi_0}\) such that \(\bm{I}^m\ket{\psi_0}=0\). Since \(\bm{I}\) is non-nilpotent, \(\bm{I}\neq 0\). Thus, by considering Eq.~(\ref{eq:A2}), we get \(\frac{1+\mathrm{e}^{\mathrm{i}\theta}}{2}\neq0\) and \(\big(\frac{1+\mathrm{e}^{\mathrm{i}\theta}}{2}\big)^m\neq0\). Also, according to that the unitary transformation preserves the normalization of the state vector, \(|(\bm{U_4}\bm{U_3})^m\ket{\psi_0}|=|\ket{\psi_0}|=1\), it is implied that \((\bm{U_4}\bm{U_3})^m\ket{\psi_0}\neq0\). We calculate \(\bm{I}^m\ket{\psi_0}\):
\begin{equation}
\begin{split}
    \label{eq:A4}
    \bm{I}^m\ket{\psi_0} &= \big(\frac{1+\mathrm{e}^{\mathrm{i}\theta}}{2}\bm{U_4}\bm{U_3}\big)^m\ket{\psi_0} \\
                          &= \big(\frac{1+\mathrm{e}^{\mathrm{i}\theta}}{2}\big)^m((\bm{U_4}\bm{U_3})^m\ket{\psi_0})\neq0,
\end{split}
\end{equation}
which is in contradiction with the assumption that \(\bm{I}^m\ket{\psi_0}=0\). Therefore, the original assumption is false, and the proof is completed.

\section{Population analysis of the ion state}
In this Appendix, we show an experimental method of measuring the population of each \(\ket{g}\otimes\ket{n}\) or \(\ket{e}\otimes\ket{n}\) state, which corresponds to $\ket{S_{1/2},m_{j}=-1/2,1/2}\otimes\ket{n}$  and $\ket{D_{5/2},m_{j}=-3/2,-1/2,1/2,3/2}\otimes\ket{n}$  in our scheme. We utilize the qudit readout method in Ref.~[\onlinecite{ringbauer2022}], combined with the phonon state readout method in Ref.~[\onlinecite{meekhof1996generation}].

In trapped ion experiments, state readout is obtained typically via fluorescence detection by driving the cycling transition between $S_{1/2}$ levels and $P_{1/2}$ levels, where scattered photons are collected via a photomultiplier tube (PMT). The $D_{5/2}$ states remain dark in such a process, while strong fluorescence occurs for both the sublevels of $S_{1/2}$. To distinguish the populations in the 6-level qudit system, electron shelving technique needs to be utilized \cite{ringbauer2022}, where only the population of the specific state to be measured is transferred to the $S_{1/2}$ manifold if it was not there, while populations of the other states are either transferred to or remain at the $D_{5/2}$ manifold. We denote the only state populated in the $S_{1/2}$ after the shelving process as $\ket{\downarrow}$. Black arrows in Fig.~\ref{fig:s} show a feasible realization of tailored resonant $\pi$-pulses with 729~nm laser fields probing various states in our scheme.
Populations of both the internal and motional states could then be detected by applying the sideband laser pulse~\cite{meekhof1996generation}, as shown with blue arrows in Fig.~\ref{fig:s}, driving a resonant sideband between $\ket{\downarrow}$ and one of the states labeled as $\ket{\uparrow}$ in the $D_{5/2}$ manifold, where the latter is ensured not populated in previous procedures. Such a process implements transition $\ket{\downarrow,n}\leftrightarrow\ket{\uparrow,n+1}$, whose Rabi frequency $\Omega_{n+1,n}$ is motional dependent. Thus, one can apply sideband drives with varied duration $\tau$, followed by fluorescence measurement. 
We subsequently measure the probability $P_{\downarrow}(\tau)$ that the ion is in the $\ket{\downarrow}$ internal state. 
The experiment is repeated many times for each value of $\tau$, covering a range of $\tau$ values. Theoretically, $P_{\downarrow}(\tau)$ can be written as:
\begin{equation}
P_{\downarrow}(\tau)=\frac{1}{2}\left(1+\sum_{n = 0}^{\infty}|C_{n}|^2\, \mathrm{e}^{-\gamma_{n}\tau}\cos(2\Omega_{n + 1,n}\tau)\right),
\end{equation}
where $|C_{n}|^2$ is the probability of the ion being in state $\ket{\downarrow}\otimes\ket{n}$. The phenomenological decay constants $\gamma_{n}$ are included to account for the decoherence that occurs during the application of the blue sideband. By numerically fitting the measured signal \(P_{\downarrow}(\tau)\), it is possible to obtain the probability distribution of vibrational state occupation $|C_{n}|^2$.
In our 6-level qudit version, we choose $\ket{\downarrow}\equiv\ket{S_{1/2},m_{j}=-1/2}, \ket{\uparrow}\equiv\ket{D_{5/2},m_{j}=-5/2}$ for $\ket{S_{1/2},m_{j}=-1/2}$ state analysis, and $\ket{\downarrow}\equiv\ket{S_{1/2},m_{j}=1/2}, \ket{\uparrow}\equiv\ket{D_{5/2},m_{j}=5/2}$ for the other five states, as shown in Fig.~\ref{fig:s}.

\nocite{*}

\bibliography{NAABC_arxiv}

\end{document}